\documentclass[12pt,preprint]{emulateapj}

\usepackage{graphicx}

\begin{document}

\title{The Distance Duality Relation from Strong Gravitational Lensing}
\author{Kai Liao$^{1}$, Zhengxiang Li$^{2}$, Shuo Cao$^{2}$, Marek Biesiada$^{2,3}$, Xiaogang Zheng$^{2}$, and Zong-Hong Zhu$^{2}$}
\affil{
$^1$ {School of Science, Wuhan University of Technology, Wuhan 430070, China.}\\
$^2$ {Department of Astronomy, Beijing Normal University, Beijing 100875, China.}\\
$^3$ {Department of Astrophysics and Cosmology, Institute
of Physics, University of Silesia, Uniwersytecka 4, 40-007 Katowice,
Poland.}
}
\email{liaokai@mail.bnu.edu.cn}

\begin{abstract}
Under very general assumptions of metric theory of spacetime, photons traveling along null geodesics and photon number conservation,
two observable concepts of cosmic distance, i.e. the angular diameter and the luminosity distances are related to each other by the so-called distance duality relation (DDR) $D^L=D^A(1+z)^2$. Observational validation of this relation is quite important because any evidence of its violation could be a signal of new physics.
In this paper we introduce a new method to test DDR based on strong gravitational lensing systems and type Ia supernovae under a flat universe.
The method itself is worth attention, because unlike previously proposed techniques, it does not depend on all other prior assumptions concerning the details of cosmological model. We tested it using a new compilation of strong lensing systems and JLA compilation of type Ia supernovae and found no evidence of DDR violation. For completeness, we also combined it with previous cluster data and showed its power on constraining DDR. It could become a promising new probe in the future in light of forthcoming massive strong lensing surveys and because of expected advances in galaxy cluster modlelling.
\end{abstract}

\section{Introduction}


There exits a very fundamental relation connecting the observed
luminosity distance $D^L$ and angular diameter distance $D^A$ at the
same redshift $z$, namely: $D^A(z)(1+z)^2/D^L(z)\equiv1$. This is the
so-called ``distance duality relation''~\citep{Etherington1933,Etherington2007} (DDR thereafter).
This reciprocity relation always holds true under the following
three general conditions:
\begin{itemize}
\item A. The spacetime is described by a metric theory of gravity;
\item B. Photons travel along null geodesics;
\item C. The number of photons is conserved.
\end{itemize}
Conditions A and B are very basic, related to the
foundations of the theory of gravity~\citep{Adler1971,Bassett2004}. Violation of C can be achieved
much easier by cosmic opacity \citep{Liao2015b}. The number of photons arriving from
standard candles, such as type Ia supernovae (SNe Ia), may be
altered simply by the absorbtion and scattering on dust on their way to the observer.
More exotic scenarios invoked in the literature, comprise
the conversion of photons (in the presence of extragalactic magnetic fields) to very
light axions~\citep{Sikivie1983,Raffelt1999}, gravitons~\citep{Chen1995},
Kaluza-Klein modes associated with
extra-dimensions~\citep{Deffayet2000} or a chameleon
field~\citep{Khoury2004,Burrage2008}. It is therefore quite important and necessary to test the
DDR because any violation of it could be a clue of new
physics. \par
Starting with the first paper on this subject \citep{Bassett2004},
many efforts have been devoted to validate the DDR with observational
data~\citep[e.g.,][]{Holanda2010,Li2011,Uzan2004,Bernardis2006,Nair2011,Holanda2012,Khedekar2011,Ellis2013,Meng2012}.
In these works, the most common and mature strategy was to
compare luminosity distances from SNe Ia (basing on Union2 and Union 2.1 SN Ia compilations
~\citep{Union2,Union2.1}) and angular diameter
distances from galaxy clusters at almost the same redshifts. The advantage of such strategy is that the number of well-measured SNe
Ia increased rapidly during recent years. However, the sample size
of $D^A$ from galaxy clusters is still limited to several dozens. What is
even more important, the angular diameter distances estimated from X-ray and SZ-studies of clusters are
sensitive to the assumptions about the hot gas density profile (simple $\beta$ or double -- $\beta$ profile) and
the underlying geometry of the cluster (spherical or elliptical). All these unknowns contribute to systematic
uncertainties which are hard to quantify. Some researchers took the opposite approach, namely assuming that DDR was valid they tried to
draw conclusions regarding galaxy clusters \citep{Holanda10, Cao2016}.
Other papers~\citep{Nair2012} used angular diameter distance data from
Baryon Acoustic Oscillations (BAO). However, there is one more
disadvantage of DDR studies performed so far. Namely most of them used distance data (both angular diameter and luminosity distances) which were
obtained under the assumption of a certain particular cosmological model (usually flat $\Lambda$CDM).
Of course DDR as a fundamental relation based only on assumptions A-C, does not depend on a cosmological model.
However, the distances inferred under the assumption of a specific cosmology may be biased, hence obscuring the issue of how
stringent the resulting constraints on possible violation of the DDR might be.
The violations of DDR found in previous studies might originate from these \citep{Holanda2010,Li2011}.

In this paper, we propose a new method of testing the DDR, which is completely free from prior assumptions concerning the details of cosmological model except the geometry, i.e., a flat FRW cosmology, while other state of the art methods furthermore depend on information like Hubble constant or dark energy ($\Lambda CDM$ or $\omega CDM$).
The idea is to consider jointly the SNe Ia and strong gravitational lensing systems. Cosmological model independence is achieved
at the prize of keeping some nuisance parameters associated with the SN Ia light-curve calibration~\citep{Yang2013} and strong lens model assumption.
Let us also stress that by its construction our test becomes independent of the SN absolute magnitude.

This paper is organized as follows: In Section2 we introduce the observations of strong lensing and SNe Ia.
Then we describe the methodology of testing DDR and present the results in Section 3.
Finally we discuss and summarize the results in Section 4.

\section{Observables and data}
Strong gravitational lensing has become an important
and powerful tool to study both background cosmology and the
structure of galaxies acting as lenses.
Being more specific: the combined measurements of stellar central velocity dispersion
$\sigma_0$ and image separations (the Einstein radius) $\theta_E$ supplemented with an assumption about
lens mass density profile, can provide us the angular diameter distance ratio $R^A(z_l,z_s)=D^A_{ls}/D^A_s$, where
$D^A_{ls}$ and $D^A_s$ are the angular diameter distances from the lens
to the source and from the observer to the source, respectively.
For example within the Singular Isothermal Sphere (SIS) model of the lens, distance ratio $R^A$ is related to observable quantities in the following way \citep{Biesiada2010}:
\begin{equation} \label{SIS}
R^A(z_l,z_s) = \frac{c^2\theta_E}{4\pi\sigma_{SIS}^2},
\end{equation}
where $c$ is the speed of light (later on in Eq.(\ref{mub}) we introduce another quantity denoted $c$ but the distinction from the speed of light should be clear from the context), $\theta_E$ is the Einstein radius and $\sigma_{SIS}$ denotes the
velocity dispersion due to lens mass distribution. Let us stress that
$\sigma_{SIS}$ need not to be exactly equal to the observed stellar
velocity dispersion $\sigma_0$~\citep{White1996}. In order to account for this we introduce a phenomenological
free parameter $f_e$ 
\citep{Kochanek1992,Ofek2003,Cao2012} defined by the relation:
$ \sigma_{SIS}=f_e\sigma_0. $
It is worth repeating that $f_e$ accounts not only for systematic errors caused by
taking the observed stellar velocity dispersion $\sigma_0$ as
 $\sigma_{SIS}$ but also for deviation of the real mass density profile
from the SIS and the effects of secondary lenses (nearby
galaxies) and line-of-sight contamination \citep{Ofek2003}.
The fractional uncertainty of $R^A$ is given by:
\begin{equation}
\delta R^A=\frac{\Delta R^A}{R^A}=\sqrt{4(\delta \sigma_{SIS})^2+(\delta \theta_E)^2}.
\end{equation}
Following the strategy adopted by SLACS team, we take the fractional
uncertainty of the Einstein radius at the level of $5\%$ for all lensing systems.
The uncertainties of $\sigma_0$ measurements are taken from the data. This approach has already been used to constrain
cosmological parameters and the evolution of elliptical galaxies
\citep{Biesiada2010,Ruff2011,Cao2012}.
 We use
a new compilation of 118 galactic strong lensing systems from Sloan
Lens ACS Survey (SLACS), BOSS Emission-Line Lens Survey (BELLS),
Lenses Structure and Dynamics Survey (LSD) and Strong Legacy Survey
(SL2S) presented in Cao et al. (2015).

The possibility of getting $R^A$ from strong lensing systems is the starting point of
our idea. In order to constrain DDR violation, $R^A$ from strong lenses will be compared with
analogous (although not exactly the same) luminosity distance ratios $R^L$ from the SNe Ia.

As a source of SN Ia data, we use JLA compilation \citep{Betoule2014}
obtained by the SDSS-II and SNLS collaborations.
It contains several low-redshift samples (z$<$0.1), all
three seasons from the SDSS-II (0.05$<$z$<$0.4), and three years
from SNLS (0.2$<$z$<$1). In total it contains 740 data points.
As was pointed out by Yang et al. (2013), data reported in the form of distance
moduli actually depend on the details of cosmological model assumed. For example, the
Union2 \citep{Union2} and Union2.1 \citep{Union2.1} compilations calibrated the SNe Ia with $wCDM$ cosmology, while the JLA used flat $\Lambda CDM$ model.
Therefore, it is inappropriate to use these distance
moduli data directly as they depend on the cosmological model assumed a priori. In our work, instead we use
the original measurements of observed B band magnitude $m_B$, the
stretch factor $x$ and color parameter $c$ in the distance modulus:
\begin{eqnarray} \label{mub}
\mu_B(z;\alpha,\beta,M_B)&=& 5logD^L(z)+25= \\ \nonumber
 &=& m_B(z)-M_B+\alpha x(z)-\beta c(z),
\end{eqnarray}
where $D^L(z)$ is in $Mpc$. Note that
$m_B, x, c$ are directly extracted from the light curve and are independent of cosmological model,
$\alpha$ and $\beta$ are nuisance parameters related to the well-known broader-brighter and bluer-brighter relationships.
The absolute B band magnitude $M_B$ is another calibrating parameter but as we will see it will cancel in the distance ratios.
Hence, the uncertainty of luminosity distance is:
\begin{equation}
\Delta D^L=(ln10/5)D^L\sqrt{(\Delta m_B)^2+\alpha^2(\Delta x)^2+\beta^2(\Delta c)^2}.
\end{equation}

\begin{figure}
  \includegraphics[width=8cm,angle=0]{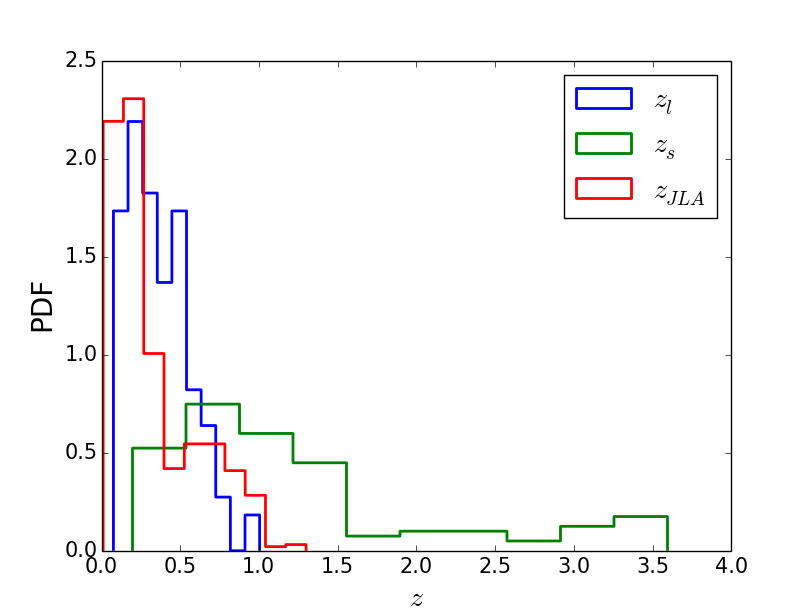} 
  \caption{Redshift distributions of lenses, sources and supernovae Ia from our 118 lensing systems and JLA sample.
  }\label{z}
\end{figure}

Redshift distributions of lenses, sources and supernovae used in our analysis are shown in Fig. \ref{z}.
As can be seen, lenses and sources overlap sufficiently with the supernovae.

\section{Method and Results}
One possible way to test DDR, is to use the following parametrization:
\begin{equation}
\frac{D^A(z)(1+z)^2}{D^L(z)}=\eta(z)\approx1+\eta_0z. \label{ddr}
\end{equation}
Standard DDR implies $\eta(z)\equiv1$. All deviations from it, which might occur at different redshifts are
encoded in the function $\eta(z)$.
Since all redshifts of the objects we use satisfy $z \leq 1$, we just take the first order term of the Taylor expansion of $\eta(z)$.

\begin{figure}
\includegraphics[width=8cm,angle=0]{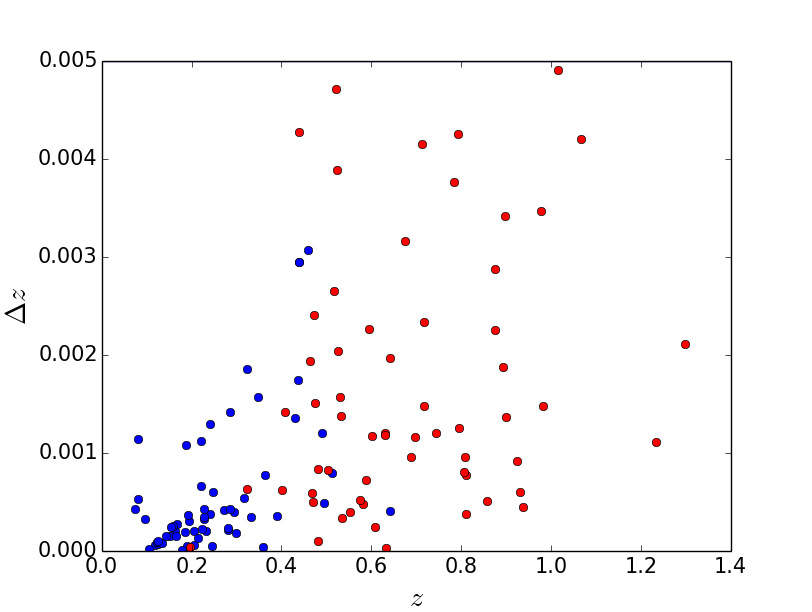}
\caption{Differences between the redshifts of SNe Ia and components of strong lensing systems: the lenses (blue points) or the sources (red points).
  The total number of selected lensing systems is 60 that all satisfy our criterion $\Delta z<0.005$
  for both lenses and sources.
  }\label{rds}
\end{figure}

The most straightforward way to test DDR would be to compare the luminosity distance and angular diameter distance at
the same redshift via Eq.(~\ref{ddr}). However, we can also test it using the distance ratios, because the strong lensing system provides us
angular diameter distance ($D^A$) ratio $R^A$. Next, by virtue of the DDR we should express it through the luminosity distance ratio $R^L$.
However the numerator $D^A_{ls}$ is the angular diameter distance from
lens to source and one can not find an observable corresponding to a similar luminosity distance from the supernova data.
Therefore, taking advantage of the fact that in flat cosmological model comoving distance $r(z) = (1+z) D^A(z)$ between lens and source is simply
$r_{ls} = r_s - r_l$, one can rewrite the $R^A$ in a simple way which contains
the ratio of $D^A_l$ and $D^A_s$:
\begin{equation}
R^A=1-\frac{(1+z_l)D^A_l}{(1+z_s)D^A_s}. \label{da}
\end{equation}
Note that in a non-flat universe, $R^A$ as a function of $D^A_l$ and $D^A_s$ can be quite complex \citep{Rasanen2015} and
it is extremely difficult to make the calculation with the covariance matrix of JLA data. For robustness and simplicity, we
only consider the flat case in this work as a first step.
Then, using the Eq.(\ref{ddr}) one can write:
\begin{eqnarray} \label{dadl}
R^A(z_l,z_s) &=& 1-\frac{D^L_l(1+z_s)(1+\eta_0z_l)}{D^L_s(1+z_l)(1+\eta_0 z_s)} = \\ \nonumber
& = &1 - R^L(z_l,z_s) p(\eta_0;z_l,z_s)
\end{eqnarray}
where
\begin{equation} \label{p}
p(\eta_0;z_l,z_s)=\frac{(1+z_s)(1+\eta_0z_l)}{(1+z_l)(1+\eta_0 z_s)}.
\end{equation}
This is the main formula used to infer the value of $\eta_0$ (capturing any possible deviations from the DDR).
It involves observables $R^A(z_l,z_s)$ and $R^L(z_l,z_s)$. Note that $R^A = D^A_{ls}/D^A_s$ and $R^L = D^L_{l}/D^L_s$ here are not exact counterparts
but they can be derived from the observed quantities. Let us also point out that, by virtue of Eq.(\ref{mub}),
$logR^L(z_l,z_s) = 0.2[ (m_B(z_l) - m_B(z_s)) + \alpha (x(z_l) - x(z_s)) - \beta( c(z_l) - c(z_s) )] $, the absolute magnitude of a fiducial SN Ia cancels out. The redshifts here have the meaning of redshifts of the supernovae located at $z_l$ and $z_s$ respectively.
Therefore, for each lensing system, we have to find two SNe Ia located at the lens and the source redshifts.
To achieve this, many approaches have been proposed. We adopt the simplest and efficient method of
Holanda et al.(2010) and Li et al. (2011), which applies the following criterion: if the
redshift difference between the lens or the source and its matched SN Ia is
smaller than 0.005, it can be ignored. To avoid correlations among
individual DDR tests, which would occur if we could select the same SN Ia pair for
different lensing systems, we adopt the procedure that if certain SN Ia pair is matched to some lensing system it
can not be used again. This procedure resulted in selecting a sample smaller than original 118 lenses.
The number of filtered lensing systems are 60 and the redshift differences are summarized
in Fig. \ref{rds}.

Next we use the PyMC Python module\footnote{{\tt http://github.com/pymc-devs/pymc}}
that implements Bayesian statistical models and fitting algorithms, including Markov chain Monte Carlo (MCMC) to calculate full posterior
PDF: $P(\eta_0, \alpha, \beta, f_e| z_l, z_s, \theta_E, \sigma_0, m_B, x, c) = {\cal L}(z_l, z_s, \theta_E, \sigma_0, m_B, x, c | \eta_0, \alpha, \beta, f_e) P(\eta_0, \alpha, \beta, f_e)$.
The likelihood is derived from the $\chi^2$ function ${\cal L} \sim exp( - \chi^2_{lensing}/2) $, where $\chi^2_{lensing}(\eta_0,\alpha,\beta,f_e)=\textbf{O}_l^T\textbf{C}_l^{-1}\textbf{O}_l, \textbf{C}$ is the
covariance matrix for the ``observational quantity" $\textbf{O}_l$. Using Eq.(\ref{dadl}), one can write each component:
\begin{equation}
O_{li}=R^A_i-1+p R^L_i.
\end{equation}

Prior probability for the parameters $P(\eta_0, \alpha, \beta, f_e)$ is the product of respective priors assumed to be uniform distributions:
$P(\eta_0)=U[-2.0,2.0], P(\alpha)=U[-0.6,0.6], P(\beta)=U[-1.0,10.0], P(f_e)=U[0.85,1.15]$. Note that for JLA data \citep{Betoule2014},
the errors include both statistic and systematic ones. The systematic error can be embodied in the full covariance matrix of all data points,
which can make the uncertainties larger. In this work, we only consider a small part of JLA data and the corresponding systematic errors. The
different feature here is the constructed observational quantities are distance ratios, therefore, we have to consider the systematic errors for
pairs.

\begin{figure}
\includegraphics[width=8cm]{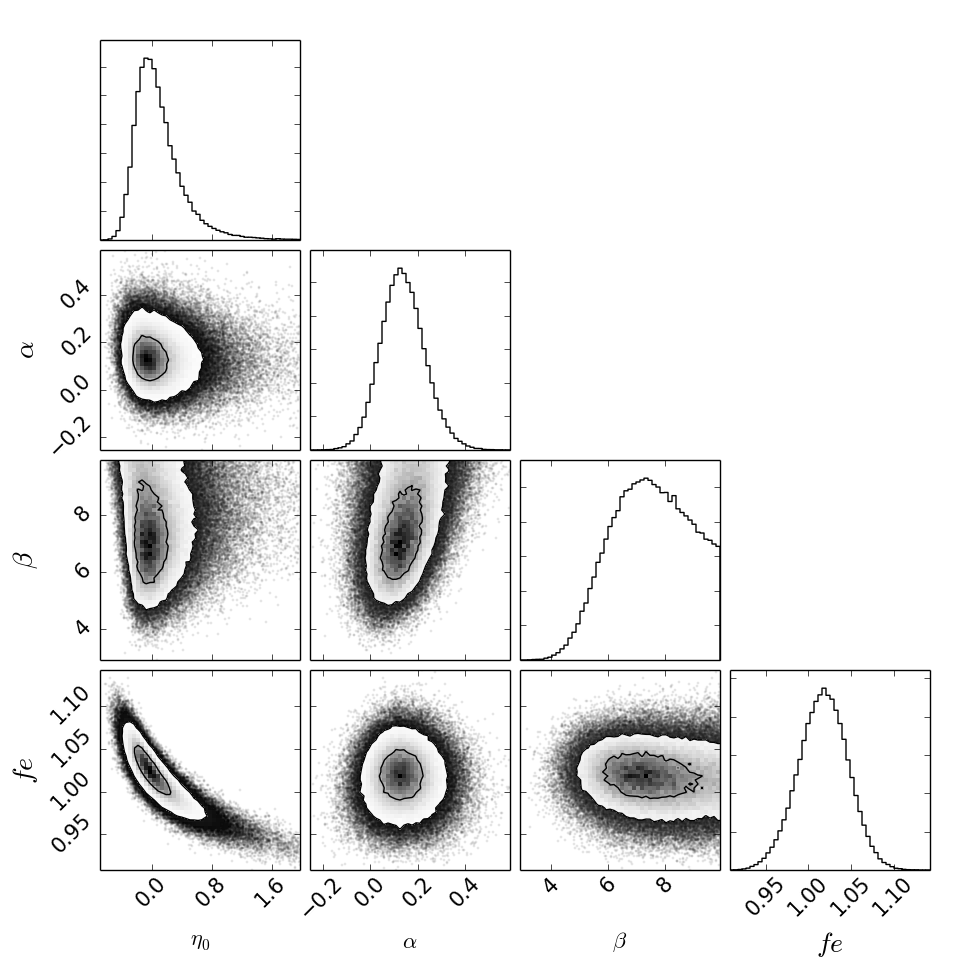}
\caption{SL: 2-D regions and 1-D marginalized distributions with 1$\sigma$ and 2$\sigma$ contours for the parameters $\eta_0, \alpha, \beta, f_e$. }
\label{result}
\end{figure}

Our results are summarized in Tab. \ref{resulttab} and shown in Fig. \ref{result}.
Constraint contours were plotted using the publicly available package ``corner.py"
\footnote{{\tt http://github.com/dfm/corner.py}}.
One can see that $\eta_0=0$ is located near the center of $1\sigma$ contour, best fitted central value being $\eta_0 = -0.005_{-0.215}^{+0.351}$
indicating that the DDR is in a very well agreement with the observations and there is no signs of its violation in light of SN Ia and strong lensing data
\footnote{We have specified the cosmological parameter $\Omega_k=0$. However, like the analysis in \cite{Liao2015b}, the independent determination of the flatness of our universe at the percent level, implies that corrections coming from deviations from flatness should be at least one order of magnitude smaller than the errors considered here.}.

\begin{figure}
\includegraphics[width=8cm]{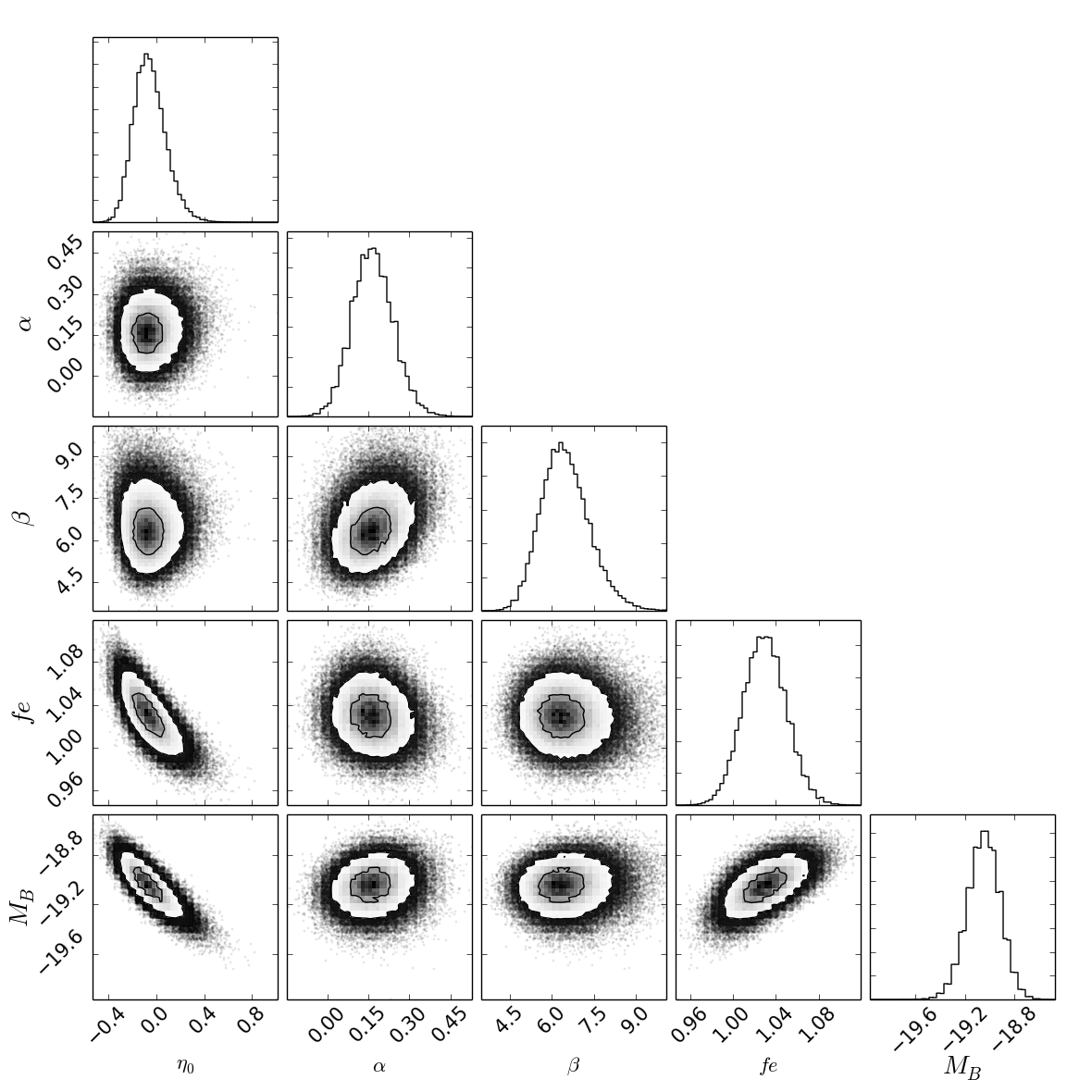}
\caption{SL+38 Clusters(S): 2-D regions and 1-D marginalized distributions with 1$\sigma$ and 2$\sigma$ contours for the parameters $\eta_0, \alpha, \beta, f_e, M_B$. }
\label{add38}
\end{figure}

\begin{figure}
\includegraphics[width=8cm]{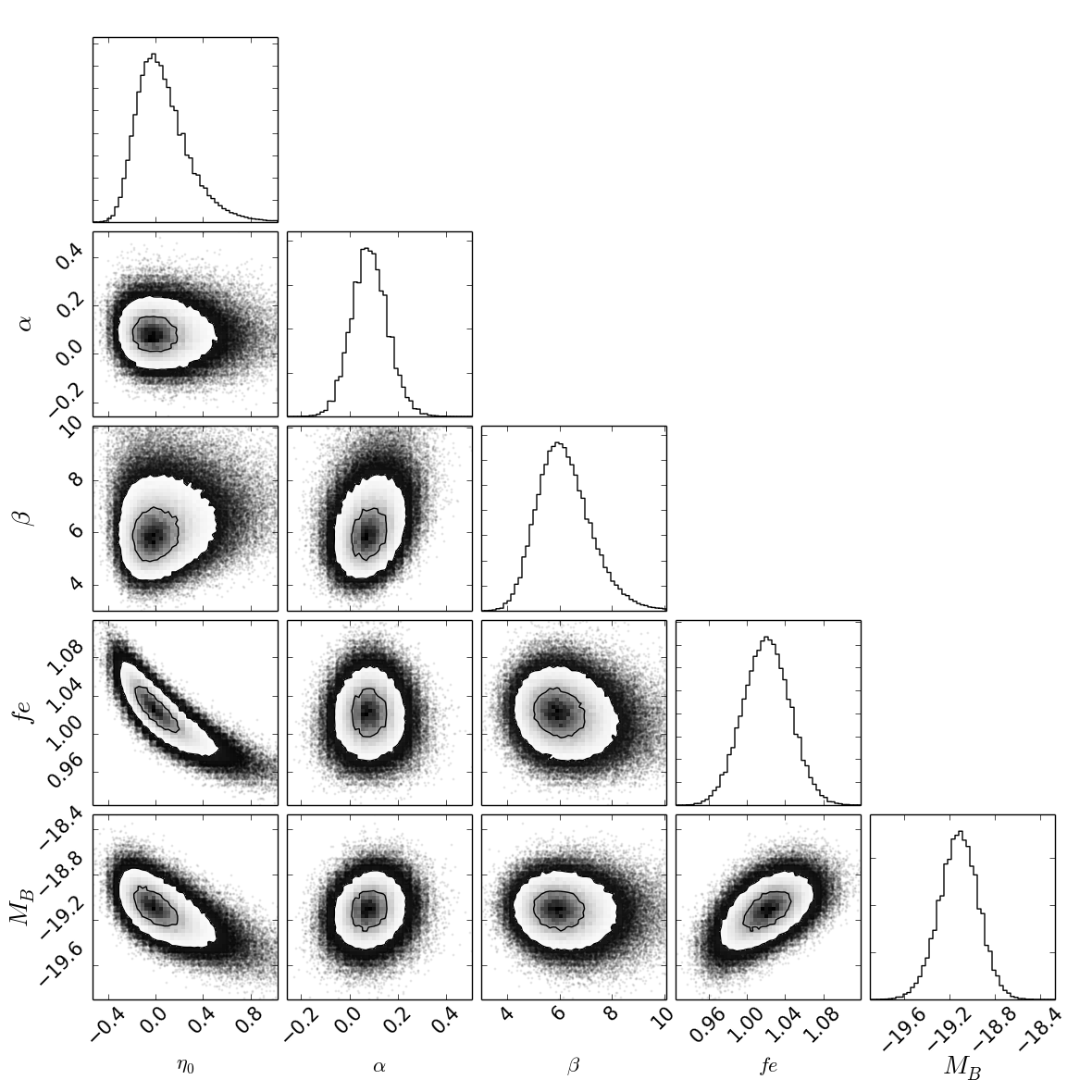}
\caption{SL+25 Clusters(E): 2-D regions and 1-D marginalized distributions with 1$\sigma$ and 2$\sigma$ contours for the parameters $\eta_0, \alpha, \beta, f_e, M_B$. }
\label{add25}
\end{figure}

In order to see how stringent the results of joint analysis would be, we have also considered galaxy clusters with angular diameter distances determined from X-ray data combined with Sunyaev-Zeldovich effect measured in these clusters. Let us stress that in principle cluster distances are also obtained in a model independent way. However, systematic uncertainties are different: for strong lensing systems they come form the ansatz of SIS augmented with $f_e$ factor, whereas for clusters they come from the model assumptions concerning gas distribution profile.
We have considered two separate data sets. First, there were 38 clusters of \citet{Bonamente2006} where spherical symmetry was assumed and the cluster plasma was assumed to be in hydrostatic equilibrium with dark matter following Navarro-Frenk-White profile. Second sample comprised 25 clusters analyzed by \citet{Fillippis2005} within elliptical model.
The analysis details are similar to \citet{Yang2013}. In particular, we used the Eq.(\ref{ddr}) to derive distance modulus to the cluster:
$\mu_{cluster}(z; \eta_0) = 5 \log{[\eta(z) D^A_{cluster}(z) (1+z)^2]} + 25 $ ($\eta(z)$ here is the inverse in Yang et al. 2013) and then we considered the chi-square function of the form
$\chi^2_{cluster}=\textbf{O}_c^T\textbf{C}_c^{-1}\textbf{O}_c$, where
\begin{equation}
O_{ci}=\mu_B(z_i;\alpha,\beta,M_B) - \mu_{cluster}(z_i; \eta_0).
\end{equation}
The covariance matrix includes the corresponding systematic errors of the SNe we use (38 or 25 points). Here, the supernova absolute magnitude $M_B$ should be included as an extra nuisance parameter.
The rest of analysis is analogous to \citep{Yang2013} but with the newest JLA sample \citep{Betoule2014} instead of Union2 \citep{Union2}.
The results are presented in Fig. \ref{add38} and Fig. \ref{add25}. One can see that joint analysis gave considerably more stringent constraints.

\begin{table*}
\large
\begin{center}
\begin{tabular}{lccc}
\hline\hline
\tableline
     &SL  & SL+38 Clusters(S) & SL+25 Clusters(E) \\
$\eta_0$ & $-0.005_{-0.215}^{+0.351}(1\sigma)_{-0.386}^{+0.907}(2\sigma)$ & $-0.051_{-0.103}^{+0.152}(1\sigma)_{-0.226}^{+0.318}(2\sigma)$ &$-0.003_{-0.169}^{+0.234}(1\sigma)_{-0.299}^{+0.549}(2\sigma)$\\
$\alpha$ &$0.129_{-0.088}^{+0.092}(1\sigma)_{-0.176}^{+0.189}(2\sigma)$  & $0.169_{-0.092}^{+0.059}(1\sigma)_{-0.160}^{+0.139}(2\sigma)$    & $0.065_{-0.055}^{+0.089}(1\sigma)_{-0.127}^{+0.165}(2\sigma)$\\
$\beta$ &$6.700_{-1.415}^{+2.017}(1\sigma)_{-1.957}^{+3.264}(2\sigma)$ & $6.173_{-0.769}^{+0.941}(1\sigma)_{-1.483}^{+1.930}(2\sigma)$ &                         $5.599_{-0.892}^{+1.150}(1\sigma)_{-1.854}^{+2.421}(2\sigma)$\\
$f_e$ &$1.022_{-0.030}^{+0.029}(1\sigma)_{-0.062}^{+0.055}(2\sigma)$  & $1.019_{-0.020}^{+0.019}(1\sigma)_{-0.041}^{+0.040}(2\sigma)$ &                      $1.024_{-0.027}^{+0.024}(1\sigma)_{-0.049}^{+0.048}(2\sigma)$\\
$M_B$ & &                                                            $-19.052_{-0.150}^{+0.111}(1\sigma)_{-0.285}^{+0.233}(2\sigma)$ &                          $-19.113_{-0.195}^{+0.156}(1\sigma)_{-0.380}^{+0.311}(2\sigma)$\\
\hline\hline
\end{tabular}\caption{  The best-fit values of parameters $\eta_0, \alpha, \beta, f_e, M_B$ with 1$\sigma$ and 2$\sigma$ uncertainties,
              for three cases: strong lensing (SL), stong lensing+clusters with spherical model (S) \citep{Bonamente2006},
              strong lensing+clusters with elliptical model (E) \cite{Fillippis2005}, respectively.
}\label{resulttab}
\end{center}
\end{table*}

\section{Discussions and Conclusions}
Validation of DDR using extragalactic observations is an important issue in modern cosmology.
Any strong enough evidence of its violation could be a signal of new physics either in the theory of gravity
or in the particle physics sector. Currently most common method to test DDR was to compare luminosity distances
from standard candles (SNe Ia) and angular diameter distances from standard rulers (galaxy clusters).
Technically such an approach demanded that some specific cosmological model be assumed together with the values of its parameters like
the Hubble constant, cosmic equation of state or matter density parameter $\Omega_m$. This resulted in additional source of uncertainties and bias.

In this paper, we proposed a new method to test distance-duality relation (DDR) which is free from such shortcomings (though it still assumes a flat universe).
The idea is to use strong lensing systems with known redshifts
to the lenses and to the sources, the Einstein rings and central velocity dispersions. These observables allow us to infer
the angular diameter distance ratios which -- by virtue of the DDR -- could be
linked to luminosity distance ratios of matched pairs of SNe Ia and hence used to test the violations of the DDR.
Using a recent compilation of 118 strong lensing (SL) systems and 740 SNe Ia from the JLA compilation we were able to select 60 matched (SL--SN) pairs to perform the test.
Our result confirmed the validity of the DDR. The method we proposed is an interesting new technique, complementary to previous methods used by the others.
Combining it with previous methods that may be improved in the future, one can effectively eliminate systematic errors from single test and break the constraint degeneracy, thus giving a more robust result.
To show this, we combined it with galaxy cluster data from spherical model and elliptical model, respectively. Despite the cluster modelling problem,
the results showed the potential power of our method. If the cluster modelling problem is solved in the future, the combination will give a cosmological-model independent result at better precision than previous cosmological-model dependent approaches \citep{Li2011,Holanda2010} can get.

The forthcoming new generation of sky surveys like the EUCLID mission, Pan-STARRS, LSST, JDEM, are estimated
to discover from thousands to tens of thousands of strong lensing systems. Even with a few percent fraction of them having spectroscopic follow-ups, this
would enlarge the SL sample considerably making the constraints on DDR violation much more stringent. Furthermore, with sufficient number of
well-measured time delays \citep{Liao2015a}, it is worth noting that our method can be extended using time-delay distance consisting of a combination of three angular diameter distances.

\section*{Acknowledgments}
This work was supported by the Ministry of Science and Technology
National Basic Science Program (Project 973) under Grants Nos.
2012CB821804 and 2014CB845806, the Strategic Priority Research
Program ``The Emergence of Cosmological Structure" of the Chinese
Academy of Sciences (No. XDB09000000), the National Natural Science
Foundation of China under Grants Nos. 11503001, 11373014 and
11073005, the Fundamental Research Funds for the Central
Universities (WUT: 2016IVA078) and Scientific Research Foundation of Beijing Normal
University, and China Postdoctoral Science Foundation under grant
No. 2015T80052.  M.B. obtained approval of foreign talent introducing
project in China and gained special fund support of foreign
knowledge introducing project.

\clearpage

\end{document}